\documentclass{article}

\usepackage{arxiv}
\pdfoutput=1
\usepackage[utf8]{inputenc} % allow utf-8 input
\usepackage[T1]{fontenc}    % use 8-bit T1 fonts
\usepackage{url}            % simple URL typesetting
\usepackage{booktabs}       % professional-quality tables
\usepackage{amsfonts}       % blackboard math symbols
\usepackage{nicefrac}       % compact symbols for 1/2, etc.
\usepackage{microtype}      % microtypography
\usepackage{lipsum}
\usepackage[table,xcdraw]{xcolor}
\usepackage{amssymb}
\usepackage{amsmath}
\usepackage{natbib}
\usepackage{multicol}
\usepackage{multirow}
\usepackage{booktabs}
\usepackage{colortbl}
\usepackage{array}
\usepackage{enumitem}
\usepackage{bbm}
\usepackage{setspace}
\usepackage{graphicx}
\usepackage{lscape}
\usepackage{graphicx}
\newcommand{\indep}{\rotatebox[origin=c]{90}{$\models$}}

\title{Generalizing Randomized Trial Findings to a Target Population using Complex Survey Population Data}

\author{
  Benjamin Ackerman\thanks{This is work in progress. Please contact Benjamin Ackerman (backer10@jhu.edu) with any questions, comments or concerns.} \\
  Department of Biostatistics\\
  Johns Hopkins Bloomberg School of Public Health\\
  Baltimore, MD 21205 \\
  \texttt{backer10@jhu.edu} \\
   \And
 Catherine R. Lesko\\
  Department of Epidemiology\\
  Johns Hopkins Bloomberg School of Public Health\\
  Baltimore, MD 21205 \\
  \texttt{} \\
   \And
  Juned Siddique \\
  Department of Preventive Medicine\\ Northwestern University\\ 
  Chicago, IL 60611\\
  \texttt{} \\
  \And
  Ryoko Susukida\thanks{This project was supported by a research award from Arnold Ventures. The content is solely the responsibility of the authors and does not necessarily represent the official views of Arnold Ventures.} \\
  Department of Mental Health\\ Johns Hopkins Bloomberg School of Public Health\\ 
  Baltimore, MD 21205\\
  \texttt{} \\
  \And
  Elizabeth A. Stuart \\
  Departments of Mental Health, Biostatistics, and\\
  Health Policy and Management \\
  Johns Hopkins Bloomberg School of Public Health\\
  Baltimore, MD 21205
}
\begin{document}
\maketitle

\begin{abstract}
Randomized trials are considered the gold standard for estimating causal effects. Trial findings are often used to inform policy and programming efforts, yet their results may not generalize well to a relevant target population due to potential differences in effect moderators between the trial and population. Statistical methods have been developed to improve generalizability by combining trials and population data, and weighting the trial to resemble the population on baseline covariates. Large-scale surveys in fields such as health and education with complex survey designs are a logical source for population data; however, there is currently no best practice for incorporating survey weights when generalizing trial findings to a complex survey. We propose and investigate ways to incorporate survey weights in this context. We examine the performance of our proposed estimator in simulations by comparing its performance to estimators that ignore the complex survey design. We then apply the methods to generalize findings from two trials - a lifestyle intervention for blood pressure reduction and a web-based intervention to treat substance use disorders - to their respective target populations using population data from complex surveys. The work highlights the importance in properly accounting for the complex survey design when generalizing trial findings to a population represented by a complex survey sample.

\end{abstract}

% keywords can be removed

\doublespacing

\section{Introduction}
\label{s:intro}

Randomized controlled trials (RCTs) are considered the gold standard for estimating the causal effect of a new treatment or intervention; however, they often suffer from poor external validity, or generalizability \citep{Shadish2002experimental, imai2008}. Evidence from RCTs is frequently used when formulating health policy and implementing new large-scale health programs, but poor generalizability may hinder policymakers' abilities to make correct policy decisions for their populations. When feasible, trial designs that strategically sample from the target population of interest to improve representativeness have been shown to also improve upon the generalizability of RCTs \citep{Peto1995, insel2006, tipton2019improved}; however, particularly in medical trials, there are many barriers to doing so, such as time, money and location. Recruitment strategies for RCTs that do not consider the ultimate target population of interest may lead to non-representative trial samples. More formally, if the trial sample differs from the target population on characteristics that moderate treatment effect, then the average treatment effect in the trial sample (SATE) will not equal the average treatment effect in the target population (PATE) \citep{cole2010generalizing}.

Several classes of post-hoc statistical methods have been developed to address concerns of generalizability once a trial has already been completed. One broad strategy uses propensity score-type methods to weight the trial so that it better resembles the target population on baseline covariates \citep{westreich2017transportability}. Note that this is similar to using propensity score weighting to estimate the average treatment effect on the treated (ATT) in non-experimental studies, where instead of fitting a model of treatment selection, a model of sample membership (i.e. trial participation vs. not) is specified. A second approach involves modeling the outcome as a flexible function of the observed covariates in the trial, and then predicting outcomes under treatment conditions in the target population. This can be done using Bayesian Additive Regression Trees (BART) \citep{hill2011bayesian, Kern2016} or Targeted Maximum Likelihood Estimation (TMLE) \citep{rudolph2014}. Lastly, doubly robust methods have been proposed, in which models are fit for both the outcome and the probability of sample membership \citep{dahabreh2019generalizing}.

The implementation of these methods requires the identification of a dataset for the target population of interest, one that contains individual-level data on all relevant treatment effect modifiers in the trial. While data availability and quality make this challenging to do \citep{stuartrhodes2017}, in practice, large nationally representative surveys collected by government agencies are often good sources of information on policy-relevant populations. For example, the National Health and Nutrition Examination Survey (NHANES) consists of a series of annual surveys that collect information on participants' demographics, socioeconomic status, dietary behaviors and health outcomes, with supplemental laboratory tests and medical examinations \citep{johnson2014national}. NHANES is designed to be representative of the non-institutionalized civilian US population across all 50 states and Washington D.C., and may therefore be a promising source of population data for implementing generalizability methods.

While surveys like NHANES may provide a wealth of information on the target population of interest, the analytic datasets on their own are themselves \emph{not} representative of the target population. These raw datasets are the result of complex survey sampling designs that systematically over-sample and under-sample certain demographic groups. Such designs may involve stratifying the target population (e.g., first by state, then by county or Census tract) and then selecting primary sampling units (e.g., households, schools, individuals) by pre-specified rates, perhaps defined by demographic categories. Some surveys implement additional levels of stratification, for example, sampling counties first and then selecting individuals within the sampled counties. Selected participants in the final sample are then assigned sampling weights inversely proportional to their probability of being selected. Additional corrections for non-response and post-stratification are also often applied \citep{valliant2013practical}. These sampling weights are typically included as a variable in the final analytic datasets, though note that not all variables used to construct the weights are always available for researchers to use. For example, sampling may occur at the zipcode level, but for confidentiality reasons, zipcode may be omitted from the final public-use dataset, while a correlated variable, such as state or region, may be included.

Given these complex survey design elements, any inferences made by weighting a trial to look like one of these survey raw datasets will generally not be accurate for the true target population, rather they will just reflect the survey sample's demographics. In other words, when using NHANES as target population data without utilizing NHANES' survey weights, one would be generalizing to the NHANES sample, \emph{not} to the non-institutionalized civilian US population. While several studies have applied these generalizability methods using population data from complex surveys, no previous work has formalized an approach for properly incorporating survey weights when doing so. 

Although the proper incorporation of complex survey design elements has not been not been addressed in the generalization context, there are some methodological similarities to be found in a limited, yet growing set of papers on using propensity score methods to estimate causal effects in non-experimental complex survey data. However, even in that context, there is no consensus on how to best use the survey weights when specifying a treatment assignment model, whether as weights or as covariates. \citet{zanutto2006comparison} argue that survey weights do not need to be used in propensity score estimation when using matching methods, so long as the survey weights are used in modeling the outcome. Through simulation studies, \citet{dugoff2014generalizing} show benefit in using the survey weights as predictors in the propensity score model, but not in using them to weight the propensity score model. \citet{ridgeway2015propensity} provide theoretical justification for weighting the propensity score model using the survey weights, and then weighting the outcome model by the resulting propensity score weights \emph{multiplied} by the survey weights. \citet{lenis2017balance} observe no difference through simulation in how the survey weights are incorporated in the propensity score model, and \citet{austin2018propensity} similarly report inconclusive findings on the optimal specification of the propensity score model. Overall, though, researchers tend to agree that ignoring survey weights altogether yields causal estimates that do not generalize to the target population in which a survey was conducted, and may produce invalid inferences when using propensity score methods. An important distinction to make is that here, we are not using the survey weights to estimate an effect \emph{within} the survey itself, rather we are using the survey as a target population to generalize \emph{to}. Other recent relevant work by \citet{yangestimation} demonstrates the benefit of a propensity score-type weighting approach when combining a non-probability sample with a companion probability sample to enhance population-level estimation. While their approach can be extended to our context by viewing RCTs as non-probability samples and surveys as population-level data, this work does not provide detailed methodological justification on the proper use of the probability-sample's survey weights.

Given this existing relevant literature, we hypothesize that it is crucial to incorporate the survey weights, which relate the survey sample back to the target population of interest, in order to correctly generalize RCT findings to the target population of interest. The rest of this paper is structured as follows: In Section \ref{s:problem}, we formally evaluate the consequences of ignoring survey weights when generalizing RCT findings to a target population on which data are available from a complex survey. We then propose an approach to estimating the population average treatment effect while incorporating survey weights in Section \ref{s:methodproposal}. In Section \ref{s:survey_simulation}, we examine our hypothesis by conducting a simulation study to investigate when the proposed approach improves our population-level inferences. We then apply the methods to two generalization examples where population data come from complex surveys in Section \ref{s:dataexamples}, and we conclude by summarizing the findings and discussing future work in Section \ref{s:conclusion}.

\section{Transporting to a Complex Survey Population Dataset}
\label{s:problem}

\subsection{Definitions and Assumptions}
\begin{figure}
 \centerline{\includegraphics[width=80mm]{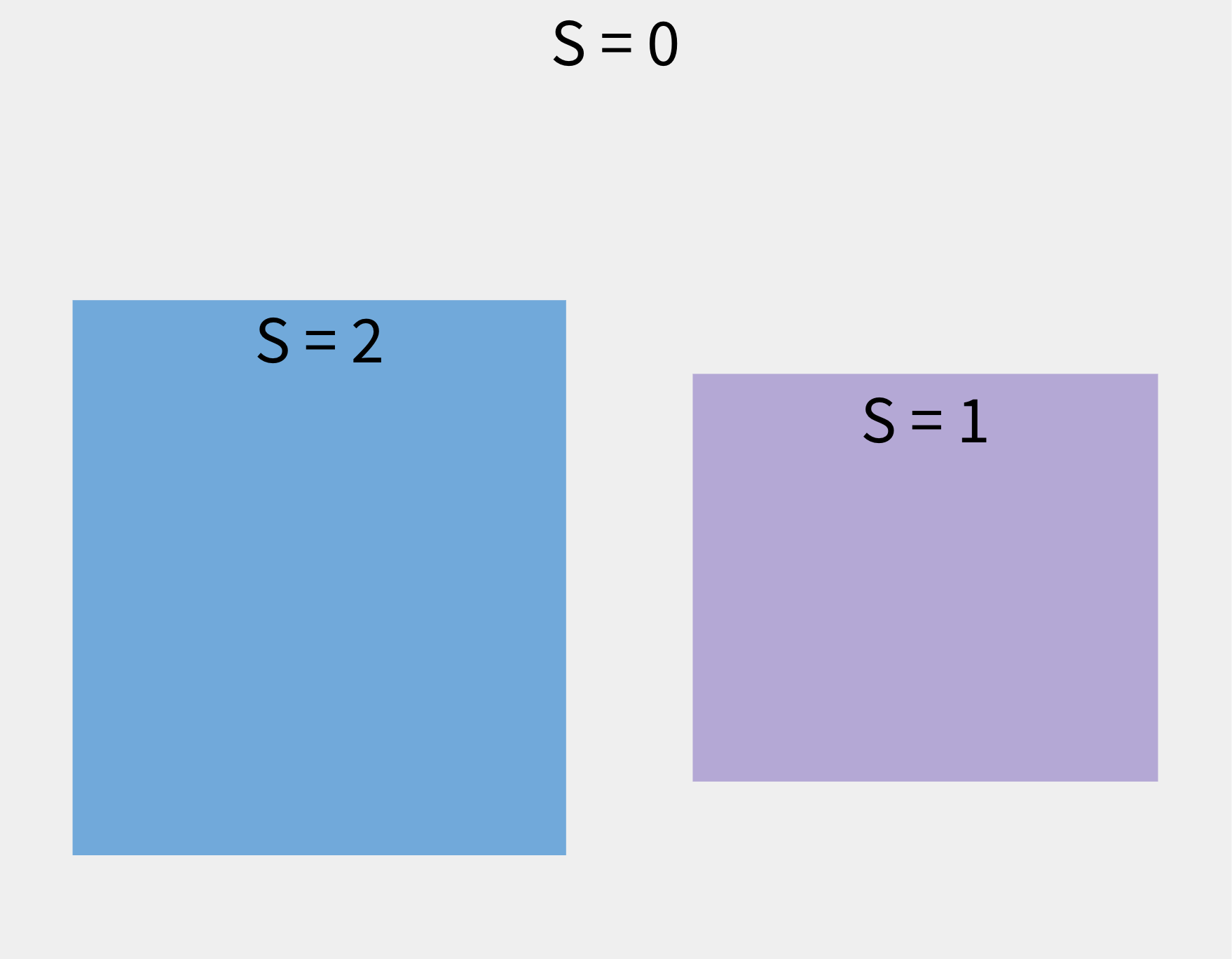}}
\caption[Scenario of how data sources relate to each other and to the target population.]{Scenario of how data sources relate to each other and to the target population. The entire grey region denotes the target population, $S=1$ denotes the RCT, $S=2$ denotes the complex survey sample, and $S=0$ denotes members of the target population not sampled into either study. Only individuals with $S=1$ or $S=2$ are observed, while data on individuals with $S=0$ are assumed unavailable. This three-level ``S" variable also assumes no overlap between trial and survey participants. This is a plausible assumption to make for policy-relevant scenarios, where the target population may be the entire US, and the study sample sizes are on the magnitudes of a few thousand.}
\label{fig:sdefinition}
\end{figure}

Suppose the goal of a randomized trial is to estimate the population average treatment effect (PATE), defined as $E[Y(1) - Y(0)]$ where $Y(a)$ is the potential outcome $Y$ under treatment $a$ ($a = 1$ denotes treatment and $a = 0$ denotes control). This expectation is defined across a well-defined target population of interest. Let $S$ denote sample membership, where $S = 1$ denotes trial membership, $S = 2$ denotes survey membership, and $S = 0$ denotes the individual is in the target population, but not the trial nor the survey sample (See Figure \ref{fig:sdefinition})\footnote{Extensions of this work could consider settings in which the trial and survey samples overlap (i.e. having two indicator variables, one for trial selection and one for survey selection).}. Here, we assume no overlap between the trial and survey samples, which is plausible for policy-relevant scenarios where the target population is the entire US and the study sample sizes are comparatively small. Additionally, let $A$ denote treatment assignment and let $X$ denote a set of pre-treatment covariates. 

Note that the population of interest is the union of all $S$ levels; however, in practice, we often do not have any data on the full population, nor do we observe outcomes for each level of $S$. Suppose all we have are data from the trial itself ($S=1$). If the RCT is a simple random sample of the target population, then we can unbiasedly estimate the PATE using the trial data alone. However, if the treatment effect in the trial is moderated by covariate $X$, \emph{and} if the distribution of $X$ differs between the trial and the target population, then the naive estimate in the trial will be a biased estimate of the PATE \citep{olsen2013external}.

In such cases, we can supplement the trial with survey data ($S = 2$) and transport the estimate of the trial to the survey to obtain an unbiased estimate of the PATE \citep{westreich2017transportability}. Note that this requires the survey data to have all $X$s related to sample selection and treatment effect heterogeneity fully observed, while treatment assignment and outcomes may be missing. Estimating the PATE by transporting the trial findings to a complex survey sample require making the following assumptions:

\begin{enumerate}
    \item[$1A$] All members of the target population have nonzero probability of being selected into the trial
    \item[$1B$] All members of the target population also have nonzero probability of being selected into the survey.
    \item[$2A$] There are no unmeasured variables associated with treatment effect and trial sample selection.
    \item[$2B$] There are also no unmeasured variables associated with treatment effect, trial sample selection \emph{and} survey sample selection.
    \item[$3$] Treatment assignment in the trial is independent of trial sample selection and the potential outcomes given the pre-treatment covariates.
    \item[$4$] The survey sample is a simple random sample of the target population (in other words, the survey is ``self-weighting")
\end{enumerate}
The plausibility of assumptions 1A, 2A and 3 have been discussed and established in previous work on generalizability. For instance, \citet{nguyen2018sensitivity} address assumption 2A by developing sensitivity analysis methods for unobserved moderators. When using population data that come from complex surveys, however, assumptions 2B and 4 must also be made. These two assumptions are under-discussed in the existing generalizability literature, and are also highly unrealistic assumptions to make given the complex survey designs of most publicly available government surveys. We now describe how biased the transported estimate will be as an estimate of the PATE when assumptions 2B and 4 are violated, and particularly, when the complex survey weights are ignored.

\subsection{Consequence of ignoring survey weights in the PATE}
Recall that the estimand of interest here is the PATE, defined as $\Delta = E[Y(1) - Y(0)]$. This estimand can be expanded upon and expressed as:
$$\Delta = E\Bigg[\frac{\mathbbm 1_{S =1} AY}{e(\emptyset)\delta^{-1}(X)} -  \frac{\mathbbm 1_{S =1}(1-A)Y}{e(\emptyset)\delta^{-1}(X)}\Bigg] $$
where $e(\emptyset) = P(A = a)$ and $\delta(X) = \frac{P(S = 2|X)}{P(S = 1|X)}\times \frac{1}{P(S = 2|X)}$. In other words, the PATE can be re-written in terms of the trial data ($S=1$) and the relationship between the trial sample and the target population ($\delta(X)$). This extends upon a result from \citet{cole2010generalizing} by recognizing that 
\begin{equation}\label{eq:weightdecomposition}
\underbrace{\Bigg( \frac{P(S = 2 | X)}{P(S = 1|X )} \Bigg)}_{\text{Transportability weights}} \times \underbrace{\Bigg(\frac{1}{P(S = 2 | X)} \Bigg)}_{\text{Survey weights}} = \Bigg(\frac{1}{P(S = 1 | X)} \Bigg)
\end{equation}
Furthermore, when $P(S = 1|X)$ cannot be estimated directly, as is often the case since RCTs are not equipped with ``trial selection weights," it can be conveniently decomposed into two estimable quantities: the inverse odds of sample vs. survey membership (transportability weights) and the inverse probability of survey sampling (survey weights).

Note that survey weights are commonly included as variables in publicly available government complex surveys. While some researchers have, in practice, incorporated survey weights when transporting from a trial to a complex survey sample, none have provided methodological details on how exactly they were used, nor have they provided any justification for their use. Without such reasoning, it is plausible that some researchers may apply current generalization methods with complex survey population data while neglecting to incorporate the survey weights. Suppose we were to \emph{ignore} the survey weights altogether. We can refer to this quantity as follows:
$$\Delta_{\text{transport}} = E[Y(1)-Y(0)|S = 2] = E\Bigg[\frac{\mathbbm 1_{S =1} AY}{e(\emptyset)\gamma^{-1}(X)} -  \frac{\mathbbm 1_{S =1}(1-A)Y}{e(\emptyset)\gamma^{-1}(X)}\Bigg]$$
Note that $\Delta_{\text{transport}}$ differs from $\Delta$ in that we substitute $\delta(X)$ for $\gamma(X)$ such that $\gamma(X) = \frac{P(S = 2|X)}{P(S = 1|X)}$. Observe that $$\Delta_{\text{transport}} = \Delta \times P(S=2|X)$$ 
In other words, if survey weights are ignored, then the estimate of $\Delta_{\text{transport}}$ will be biased as an estimate for $\Delta$, the PATE, by a factor of $P(S=2|X)$, or the probability of being sampled for the survey given covariates $X$. Note that $\Delta_{\text{transport}}$ will only be equal to $\Delta$ when $P(S = 2|X) = 1$, or when the survey is either a simple random sample of the population, \emph{or} it is the entire finite target population. 

\section{Estimating the PATE, $\Delta$}
\label{s:methodproposal}
We now discuss three different potential estimators to estimate $\Delta$, the last of which will incorporate the complex survey weights. First, if we were to use the trial data alone ($S = 1$) to estimate $\Delta$, we could use the following naive estimator:
$$\hat \Delta_{\text{naive}} = \frac{\sum_{i} \mathbbm 1_{S_i = 1}A_i Y_i}{\sum_i \mathbbm 1_{S_i = 1}A_i} - \frac{\sum_{i} \mathbbm 1_{S_i = 1}(1 -A_i) Y_i}{\sum_i \mathbbm 1_{S_i = 1}(1-A_i)}$$
However, recall from Section \ref{s:problem} that $\hat \Delta_{\text{naive}}$ will be a biased estimate of the PATE if the treatment effect is moderated by a pre-treatment covariate and sample selection also depends on that covariate. To improve upon this, we can transport the estimate to the survey ($S = 2$) with the following inverse-odds of sample membership weighted estimator:
$$\hat \Delta_{\text{transport}} = \frac{\sum_{i} \mathbbm 1_{S_i = 1}A_i Y_i \hat \gamma_i}{\sum_i \mathbbm 1_{S_i = 1}A_i \hat \gamma_i} - \frac{\sum_{i} \mathbbm 1_{S_i = 1}(1 -A_i) Y_i \hat \gamma_i}{\sum_i \mathbbm 1_{S_i = 1} (1-A_i) \hat \gamma_i}$$
where $\hat \gamma_i = \gamma(X_i,\hat \beta)$ and $\gamma(X,\beta) = \frac{P(S = 2|X)}{P(S = 1 |X)}$. Note that $\hat \gamma_i(X_i, \hat \beta)$ can be estimated parametrically by fitting a logistic regression model of sample membership (trial vs. survey) conditional on pre-treatment observables in a dataset in which the trial and survey data have been concatenated. While $\hat \Delta_{\text{transport}}$ may be unbiased for $\Delta_{\text{transport}}$ \citep{westreich2017transportability}, it will still be a biased estimate of the PATE, $\Delta$, if the complex survey is not ``self-weighting." We therefore propose a modified version of this estimator, one that incorporates the complex survey weights relating the survey sample ($S=2$) to the target population:
$$\hat \Delta_{\text{svy.wtd}} = \frac{\sum_{i} \mathbbm 1_{S_i = 1}A_i Y_i \hat \delta_i}{\sum_i \mathbbm 1_{S_i = 1}A_i \hat \delta_i} - \frac{\sum_{i} \mathbbm 1_{S_i = 1}(1 -A_i) Y_i \hat \delta_i}{\sum_i \mathbbm 1_{S_i = 1} (1-A_i) \hat \delta_i}$$
where $\hat \delta_i = \delta(X_i,\hat \beta)$ and $\delta(X,\beta) = \frac{P(S = 2|X)}{P(S = 1 |X)} \times \frac{1}{P(S = 2|X)}$. Here, $\hat \delta_i$ can be estimated parametrically by fitting a model for $\frac{P(S = 2|X)}{P(S = 1 |X)}$, and multiplying the resulting estimated transportability weights by the survey weights. If all related covariates are observed and accounted for, then this estimator is unbiased for the PATE, directly following a result from \citet{buchanan2018generalizing} by applying the equality in Equation \ref{eq:weightdecomposition}. We will now present a simple example to compare each of these estimators when weighting a trial to a target population based on a single covariate. 

\subsection{Toy Example}
In order to highlight the consequences of ignoring survey weights when estimating the PATE, consider the scenario in Table \ref{tab:toyexample}. Suppose that in the true target population of interest, 50\% of people are above the age of 40, while the other 50\% are 40 or younger. Suppose data on the full target population are not available, but a survey is conducted among the target population members, where 200 individuals over the age of 40 and 300 individuals who are 40 or younger are sampled. In order for the survey to be representative of the target population according to dichotomous age, survey weights are constructed as the inverse probability of being sampled into the survey given age category. The older category individuals are given a weight of $\frac{5}{2}$ while the younger category individuals are assigned a weight of $\frac{5}{3}$. In doing so, older survey participants receive greater weight than younger ones to reflect that older individuals are undersampled in the survey.

\begin{table}[h]
\centering
\caption{Toy example of a population (not observed), a survey sampled from the population with weights to reflect the population demographics distribution, and a trial sampled from the population (by convenience sampling)}
\label{tab:toyexample}
\begin{tabular}{l|c|c|c|c|}
\cline{2-5}
                                  &  $E[Y(1)-Y(0)|X]$ &\cellcolor[HTML]{C0C0C0}Target pop & Survey & RCT \\ \hline
\multicolumn{1}{|l|}{age > 40}     & 2& \cellcolor[HTML]{C0C0C0} 500 & 200    & 100 \\ \hline
\multicolumn{1}{|l|}{age $\leq$ 40} & 4&\cellcolor[HTML]{C0C0C0}  500 & 300    & 50  \\ \hline
\end{tabular}
\end{table}

Next, suppose a randomized trial is conducted among a convenience sample from the population, and among the recruited participants, $\frac{2}{3}$ of them are 40 years or older. Additionally, suppose that the treatment effect is truly moderated by age, where younger participants experience twice the average effect as older participants. Observe that while older members are \emph{undersampled} in the survey, they are \emph{oversampled} in the trial, and since age moderates treatment effect and differs between the trial and population, the RCT findings will not generalize well to this target population.

First, the true PATE can be calculated by averaging over the stratum-specific treatment effects in the target population: 
$$\Delta = \sum_x E[Y(1)-Y(0)|X=x]P(X=x) = 2 \times 0.5 + 4 \times 0.5 = 3$$
Next, the naive trial estimator for the PATE can be estimated as follows:
$$\hat \Delta_{\text{naive}} = \sum_{x} E[Y(1)-Y(0)|X=x]P(X=x|S=1) = 2\times\frac{2}{3} + 4\times\frac{1}{3} = 2.67$$
As expected, the naive estimate is an underestimate of the PATE because the trial oversampled older participants, while the treatment has a stronger effect for younger participants. If we apply the standard transportability weighting methods using the survey as the target population dataset, and if we \emph{ignore the survey weights}, we would weight trial members by the inverse odds of trial participation conditional on their age category. Older trial participants would be given a weight of $\frac{200}{100} = 2$, and younger trial participants would be given a weight of $\frac{300}{50} = 6$. We would therefore estimate the transported estimate as follows:
\begin{align*}
\hat \Delta_{\text{transport}} &= \frac{\sum_{x} E[Y(1)-Y(0)|X=x]P(X=x|S=1)\frac{P(S = 2|X=x)}{P(S=1|X=x)}}{\sum_x P(X=x|S=1)\frac{P(S = 2|X=x)}{P(S=1|X=x)}} \\
&= \frac{2\times\frac{2}{3}\times 2 + 4\times\frac{1}{3}\times 6}{\frac{2}{3}\times 2 + \frac{1}{3}\times 6} = 3.2
\end{align*}
As a result, the estimate is unbiased for the ATE in the \emph{survey}; however, it is still biased as an estimate of the PATE. Our inferences here reflect that \emph{older} participants are oversampled in the survey, and so in this case we are overestimating the true PATE. Finally, if we utilize the survey weights by \emph{multiplying} the inverse odds transportability weights by the inverse probability of survey selection, we would obtain weights of $\frac{200}{100}\times \frac{500}{200} = 5$ and $\frac{300}{50}\times \frac{500}{300} = 10$ for the older and younger trial participants, respectively, thereby accurately weighting them to the \emph{target population} age distribution. We would estimate the PATE using this approach as follows:
\begin{align*} 
\hat \Delta_{\text{svy.wtd}} &= \frac{\sum_{x} E[Y(1)-Y(0)|X=x]P(X=x|S=1)\frac{P(S = 2|X=x)}{P(S=1|X=x)}\frac{1}{P(S = 2|X=x)}}{\sum_x P(X=x|S=1)\frac{P(S = 2|X=x)}{P(S=1|X=x)}\frac{1}{P(S=2|X=x)}}\\
&= \frac{2\times\frac{2}{3}\times 5 + 4\times\frac{1}{3}\times 10}{\frac{2}{3}\times 5 + \frac{1}{3}\times 10} = 3
\end{align*}
Observe that our estimate of the PATE is now unbiased, as we are accounting for the fact that our survey is not ``self-weighting" and the survey weights must therefore be used to make inferences relevant to the true target population of interest.

\subsection{Estimating $\hat \Delta_{\text{svy.wtd}}$ with a weighted sample membership model}
When accounting for a small set of covariates, such as in the example above, one can directly construct and multiply the transportability weights by the survey weights. When using a survey equipped with pre-estimated survey weights, though, this is not plausible.  We therefore propose a two-stage weighting approach, where we first weight the sample membership model using the survey weights before constructing the inverse odds transportability weights. This is equivalent to the multiplication of weights in the simple approach above, because by weighting survey participants in the sample membership model, we are recognizing that each participant represents a particular number of individuals in the true target population. For example, if a survey participant has a probability of survey selection of 0.02, the corresponding weight of $\frac{1}{0.02} = 50$ suggests that the individual should count for 50 people in the population when estimating population effects with the survey. Weighting the survey participants in the sample membership model allows us to therefore compare the trial demographics to the target population, and \emph{not} to the survey sample.

The first step entails fitting a weighted logistic regression model of sample membership using a pseudo-likelihood approach \citep{pfeffermann1993role}, where trial participants are assigned a weight of $1$ while survey participants are assigned weights equal to their inverse probability of survey selection. Again, these weights are typically included in complex survey datasets and are meant to be used in analyses to relate the survey back to the target population of interest. The second step entails using the predicted probabilities from the sample membership model, $\hat e_i$ to construct the inverse odds weights ($\hat \delta_i$) that are used to estimate $\hat\Delta_{\text{svy.wtd}}$, where trial participants are assigned a weight of $\frac{1-\hat e_i}{\hat e_i}$ and survey participants are assigned a weight of $0$. It is important to note that, in theory, this approach will yield an unbiased estimate of $\Delta$ only when we account for \emph{all} covariates that impact treatment effect heterogeneity and sample selection. However, it may be the case that certain variables used to construct the survey weights may not be available in the trial data, or even in the survey dataset itself. In other words, if a moderator is accounted for in the survey weights, but cannot be directly accounted for in the transportability weights as well, the PATE estimate may still be biased. With this in mind, we now describe a simulation study to compare our two-stage weighting approach to the standard transported estimator and the naive trial sample estimator.

\section{Simulation}
\label{s:survey_simulation}
We conducted a simulation study to assess the performance of the two-stage weighting approach described in the Section \ref{s:methodproposal}. We first simulated a finite population of size $N = 1000000$ with six covariates using the multivariate Normal distribution with mean vector 0, and a variance-covariance matrix where each variable had variance 1, and pair-wise correlation (i.e. $X_1$ and $X_2$, $X_3$ and $X_4$, $X_5$ and $X_6$) of $\rho$. We paired the covariates in this way and varied $\rho$ to look at scenarios where a covariate related to the sample selection mechanisms was not available in the analytic datasets, but a variable correlated with the missing covariate was available for use in its place. For example, survey participants may be sampled proportional to their zipcode, but the survey dataset might only include state as a geographic indicator for privacy purposes.

We then assigned probabilities of survey selection and trial selection to everyone in the population according to the following two models:

$$P(S_i = 1) = \text{expit}[\gamma_1(X_{1i} + X_{2i} + 2X_{3i} + 0X_{4i} + X_{5i} + 0X_{6i})]$$
$$P(S_i = 2) = \text{expit}[\gamma_2(2X_{1i} + 0X_{2i} + X_{3i} + X_{4i} + X_{5i} + 0X_{6i})]$$

We used scaling parameters $\gamma_1$ and $\gamma_2$ to control the magnitude of difference between the two samples and the population, while fixing the relative impacts of each covariate for each model. As the scaling parameters increase, the samples differ more greatly from the target population. The coefficients for the covariates were set to different values in each model to ensure that the sampling mechanisms for the trial and the survey differed from one another. Next, we generated potential outcomes for the entire population as $Y(0) \sim N(0,1)$ and $Y(1) \sim N(2 + \gamma_3[\sum_{i=1}^{6}X_i],1)$, such that the $PATE = 2$, and the $\gamma_3$ scaling parameter controlled the amount of treatment effect heterogeneity due to the covariates. Note that when $\gamma_3 = 0$ (no treatment effect heterogeneity), all of the PATE estimates should be unbiased.

In each simulation run, we then randomly sampled approximately 600 trial participants and approximately 4000 survey participants according to each individual's respective selection probabilities. In order to do this, we scaled each individual's originally generated $P(S_i = 1)$ by 0.0006 and their $P(S_i = 2)$ by 0.004, and estimated their probability of \emph{not} being selected into either study as $P(S_i = 0) = 1 - P(S_i = 1) - P(S_i = 2)$. This type of scaling combined the specified selection models with the desired sampling proportions from the population, and allowed us to then randomly generate an $S$ of $0,1 \text{ or } 2$ for each individual using a multinomial distribution. For the survey participants ($S = 2$), we retained their $P(S_i = 2)$ as their known survey sampling probabilities to construct survey weights. For the trial participants, we generated a randomized binary treatment variable $A$, as well as the observed outcome $Y = A\times Y(1) + (1-A) \times Y(0)$.

Once the trial and survey data were simulated, we estimated the PATE in the following three ways: 1) Naive trial estimator ($\hat\Delta_{\text{naive}}$), 2) transported estimator (trial-to-survey) while ignoring the survey weights ($\hat\Delta_{\text{transport}}$) and 3) transported estimator (trial-to-survey) using the survey weights to fit a weighted sample membership model ($\hat\Delta_{\text{svy.wtd}}$). For the two transportability estimators, we predicted the probabilities of sample membership by fitting models with logistic regression, generalized boosted models (GBM) and the Super Learner. GBM is a flexible, iterative algorithm that has been demonstrated to perform well when used to estimate propensity scores in non-experimental studies, capturing nonlinear relationships between covariates and treatment assignment \citep{mccaffrey2004propensity}. The Super Learner fits a series of models based on a user-specified library of methods, combining the resulting estimates such that the overall performance is no worse than the performance of the best individual method \citep{van2007super}. We considered two Super Learner libraries \citep{luedtke2016super, moodie2017treatment}, and fit each of the estimators described above using the `WeightIt' package in R \citep{WeightIt, Rprogramming}.

Lastly, in order to investigate scenarios where variables used to construct survey weights are omitted from the survey dataset, we fit the sample membership model by using all of the covariates, by omitting $X_1$, and by omitting $X_1$, $X_3$ \emph{and} $X_5$. To evaluate the performance of each method, we calculated the bias and the empirical 95\% coverage of each estimator, using $PATE = 2$ as the truth. Standard error estimates were obtained by using a standard sandwich variance estimator. We also calculated coverage for a subset of simulation scenarios using a stratified double bootstrapping approach, in which both the trial and the survey were sampled with replacement upon each bootstrap run. Strata for survey re-sampling were defined by survey probability deciles, and survey weights in each bootstrap sample were adjusted according to \citet{valliant2013practical} (see Appendix for details). The results presented in the next section are averaged over 1000 simulation runs, and are stable to the 2nd decimal place across different seeds.

\subsection{Simulation Results}

We now present the findings of the simulation study. Given the number of parameters to vary, we present figures where $\rho = 0.3$ (pairwise X correlation) and $\gamma_3 = 0.3$ (treatment effect heterogeneity).  Note though that as expected, when $\gamma_3 = 0$, all estimators were unbiased for the PATE across all scenarios.

\begin{figure}
 \centerline{\includegraphics[width=145mm]{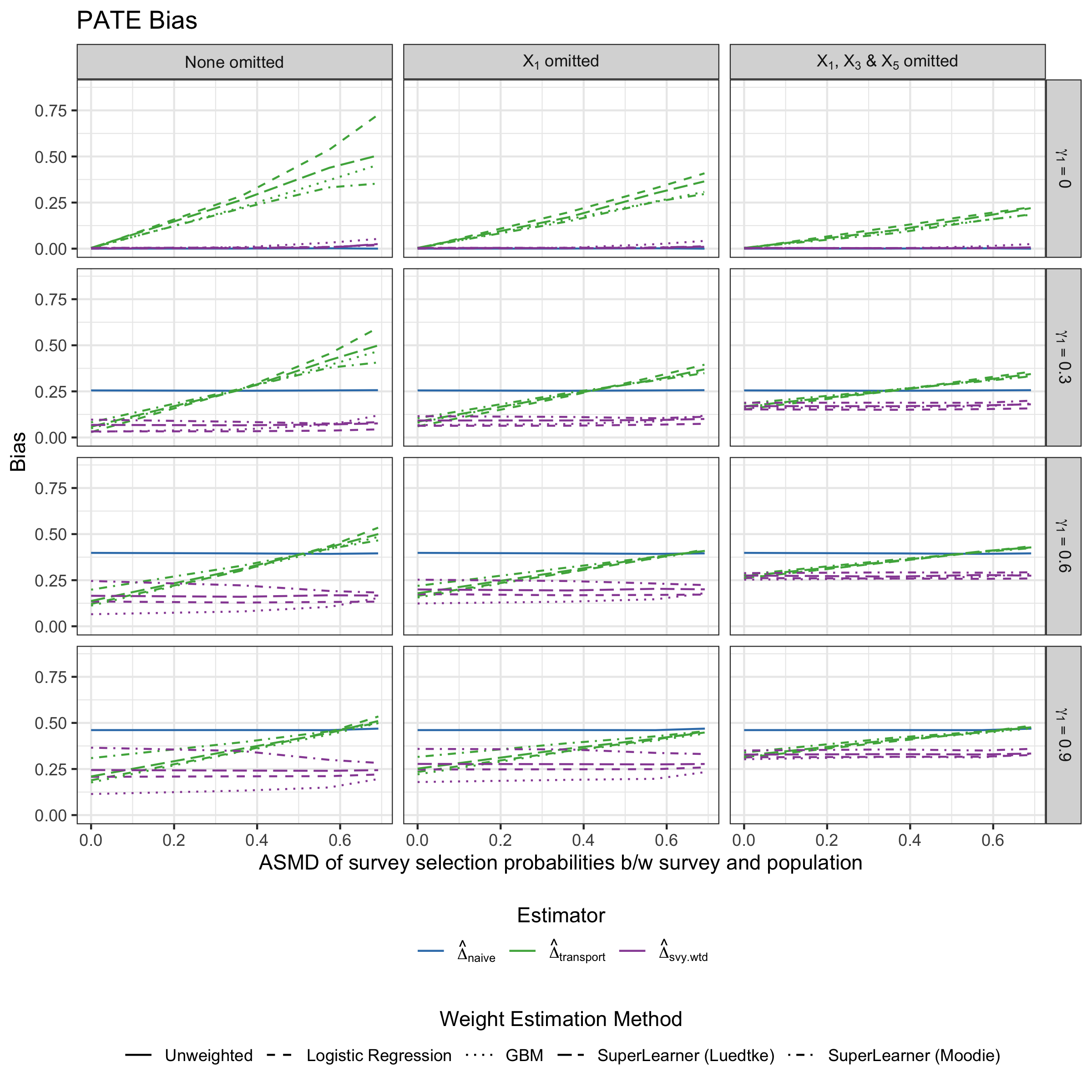}}
\caption[Bias of estimating the PATE by weighting method.]{Bias of estimating the PATE by weighting method. Each column represents a different scenario of missing a variable used to calculate survey weights in the analytic survey dataset. From top to bottom row, the $\gamma_1$ ``scale" parameter for how much the trial differs from the population by the $X$s increases. The different colors represent the different weighting approaches: Naive trial estimate (blue), transported estimate ignoring the survey weights (green), and transported estimate using the survey weights (purple). This figure appears in color in the electronic version of this article.}
\label{fig:biasresults}
\end{figure}

\begin{figure}
 \centerline{\includegraphics[width=105mm]{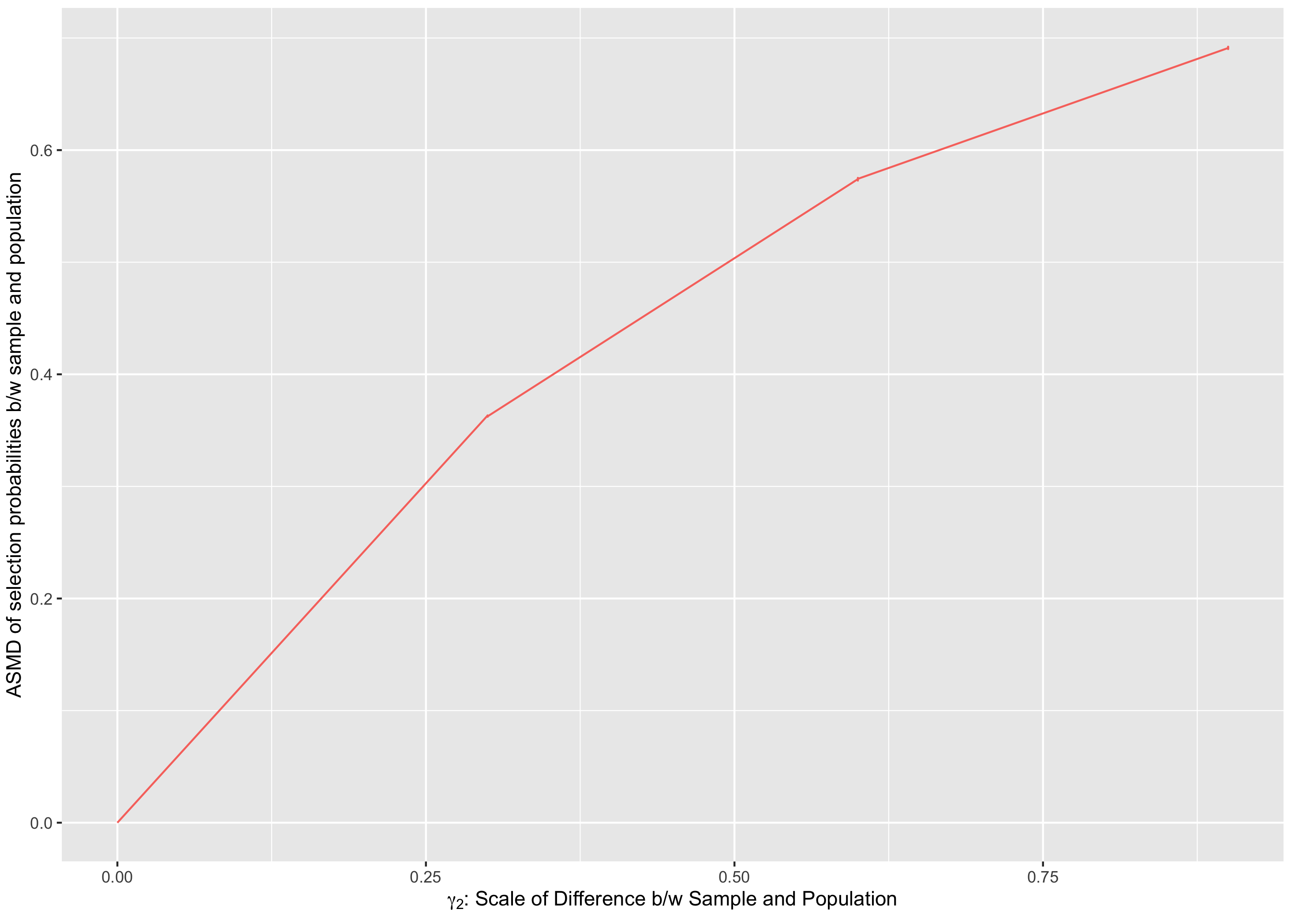}}
\caption{Relationship between $\gamma_2$, the scaling parameter for survey selection, and the ASMD of survey selection probabilities between the survey sample and the target population.}
\label{fig:asmdtoscale}
\end{figure}

When $\rho = 0.3$ and $\gamma_3 = 0.3$, Figure \ref{fig:biasresults} shows the bias of the three PATE estimators across simulation scenarios. Each column signifies a different setting regarding which variables are omitted from the sample membership model: on the left, all variables are included, and on the far right, $X_1$, $X_3$ and $X_5$ are all missing from the analytic datasets, but they were used to calculate the survey weights in the survey. Within each plot, the x-axis depicts the absolute standardized mean difference (ASMD) of the predicted probability of survey sampling between the survey sample and the target population (see Figure \ref{fig:asmdtoscale} for the relationship between $\gamma_2$ and ASMD). In other words, moving from left to right along the x-axis, the survey sample becomes increasingly different from the target population on baseline covariates. The top row depicts when $\gamma_1 = 0$, or when the trial is a simple random sample from the target population. Notice that the naive estimate is unbiased, as is the transported estimate that uses the survey weights. However, when the ATE is transported from a representative trial to a \emph{non}-representative survey and the survey weights are not used, the transported ATE estimate becomes increasingly biased as the survey becomes less representative of the population. This suggests that if findings from a trial are already generalizable, yet researchers implement transportability weighting methods without survey weights to a complex survey that is not representative of the target population, then they may actually obtain a more biased PATE estimate than had they not transported at all.

As the trial differs more greatly from the target population (moving down the rows, $\gamma_1 = 0.3$ to $\gamma_1 = 0.9$), the naive trial estimate becomes increasingly biased as expected. When the survey is slightly different from the target population of interest, the transported estimate that ignores survey weights is \emph{less} biased then the naive estimate. However, once the survey differs enough from the target population, ignoring the survey weights when transporting yields similar bias to the naive estimator, and in some cases, even greater bias. On the other hand, the transported estimate that uses the survey weights to fit a weighted sample membership model is uniformly less biased than the other estimators across all scenarios. In other words, it seems as though using the survey weights in the sample membership model can help prevent any additional bias introduced from the survey not being a simple random sample from the population.

Between the different methods used to fit the trial membership model, there is little to no difference in terms of PATE bias for $\hat \Delta_{\text{svy.wtd}}$, except for when the trial differs greatly from the target population. In such cases, predicting the probability of trial membership using GBM appears to yield the least biased ATE estimates, with notable differences in performance between the two SuperLearner libraries considered.

Next, observe that the transported estimators perform best when the selection model is fit using all covariates used to calculate the survey weights. However, when one of the variables influencing survey selection (i.e. $X_1$) is not available in the survey dataset, the bias of the transported estimators increases, and continues to increase as fewer variables impacting survey selection are included in the analytic dataset. However, as the pairwise correlation of the missing and non-missing covariates increases, the bias decreases. In other words, and not surprisingly, if $X_1$ is unavailable to use in the sample membership model, but $X_2$ is available, the more $X_2$ and $X_1$ are correlated, the less it matters that $X_1$ is missing in terms of bias.

\begin{figure}
 \centerline{\includegraphics[width=145mm]{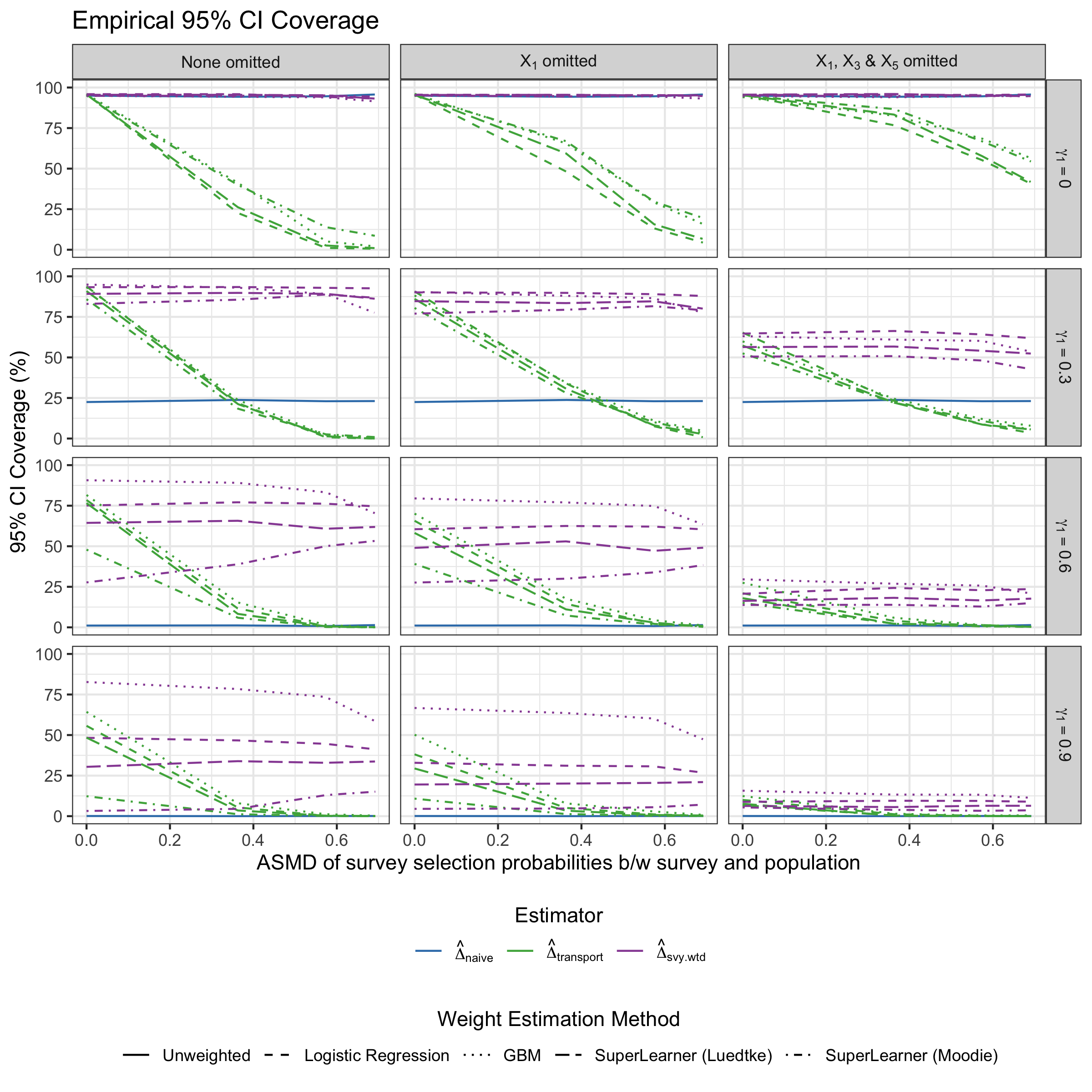}}
\caption[Empirical 95\% coverage of the PATE estimates by weighting method.]{Empirical 95\% coverage of the PATE estimates by weighting method. Each column represents a different scenario of missing a variable used to calculate survey weights in the analytic survey dataset. From top to bottom row, the $\gamma_1$ ``scale" parameter for how much the trial differs from the population by the $X$s increases. The different colors represent the different weighting approaches: Naive trial estimate (blue), transported estimate ignoring the survey weights (green), and transported estimate using the survey weights (purple). This figure appears in color in the electronic version of this article.}
\label{fig:coverageresults}
\end{figure}

Figure \ref{fig:coverageresults} shows the empirical 95\% coverage of the three estimators across simulation scenarios. Note that a standard sandwich variance estimator was used for all weighting approaches here, and results were fairly similar when using the double bootstrap approach as well (see Appendix for results). Across the top row, where the trial is representative of the target population, the coverage of the naive estimator is around 95\%, as expected (as is the coverage of the transported estimator using the survey weights). However, the coverage of the transported estimator that ignores the survey weights rapidly decreases as the survey becomes less representative of the population. Note that this corresponds to when the bias of the transported estimator without survey weights increases as well. As the trial becomes more different from the population, the empirical coverage of the naive estimator drops to zero. The transported estimator that incorporates the survey weights maintains much better coverage than the estimator that ignores the survey weights as the survey becomes less representative of the population. Also, when the trial differs substantially from the target population, the $\hat \Delta_{\text{svy.wtd}}$ estimate using GBM to fit the trial membership model results in the best coverage of the $\hat \Delta_{\text{svy.wtd}}$ estimates. The variability in the performance of $\hat \Delta_{\text{svy.wtd}}$ using the two Super Learner libraries is also notable, highlighting the method's sensitivity to library choice. Lastly, note that the transported estimator performs best when all variables included in the survey selection model are available in the survey dataset, and the empirical coverage declines as fewer of those variables become available for use in the sample membership transportability model (as $\rho$ increases, the empirical coverage improves slightly across scenarios as well).

\section{Data Examples}
\label{s:dataexamples}
We now present two applications of these methods to generalizing trial findings to well-defined target populations. First, we generalize findings from PREMIER, a lifestyle intervention trial for reducing blood pressure, to the National Health and Nutrition Examination Survey (NHANES). Next, we generalize results from CTN-0044, a trial examining the use of a web-based intervention for substance use disorder (SUD) treatment, to the National Survey on Drug Use and Health (NSDUH). In both examples, data on the respective target populations come from publicly available government surveys with complex survey sampling designs, where each participant is assigned a survey weight indicative of the number of individuals they represent in the target population. For each example, we illustrate the importance of utilizing the survey weights when comparing covariate distributions between the trial and survey, and demonstrate how the use of the survey weights affects PATE estimation. Given the simulation findings, we fit the sample membership model using GBM in both examples. 

\subsection{Lifestyle Intervention Trial for Blood Pressure Reduction}
\label{s:datanutrition}
PREMIER was a multi-center randomized trial in which 810 participants were randomized to either one of two behavioral interventions, comprised of a mix of diet and exercise recommendations, or to standard care. The primary goal of the trial was to study the effect of these lifestyle interventions on blood pressure reduction. The original report on the trial found evidence supporting the interventions' effectiveness on blood pressure reduction, and concluded that ``results from PREMIER should influence policy pertaining to implementation of lifestyle modification in the contemporary management of patients with above-optimal blood pressure through stage 1 hypertension" \citep{svetkey2003premier}. For illustrative purposes, we combine the two intervention arms into a single ``lifestyle intervention treatment" group, and select our outcome of interest as change in systolic blood pressure (SBP) between baseline and 6-month study followup.

We will now further investigate how these findings generalize to a potentially policy-relevant target population. To do so, we use population data from NHANES, a national survey funded by the Centers for Disease Control and Prevention (CDC) with extensive measures on participants' dietary behaviors and health outcomes. Using a complex and multistage probability-based sampling design, NHANES participants are carefully sampled according to sex, age, race, ethnicity and income, resulting in a sample that is representative of the entire non-institutionalized civilian US population \citep{NHANES2003}. To define the target population of interest, we subset the NHANES sample to individuals who are 25 years of age or older with BMI between 18.5 and 40 (due to PREMIER inclusion criteria). To better determine how PREMIER findings may impact a population of adults with ``above-optimal blood pressure," we further limit the NHANES sample to individuals with either SBP greater than or equal to 120 \emph{or} diastolic blood pressure (DBP) greater or equal to 80. This results in a sample size of 2180 representing a population of over 85 million US adults.

\begin{figure}
 \centerline{\includegraphics[width=145mm]{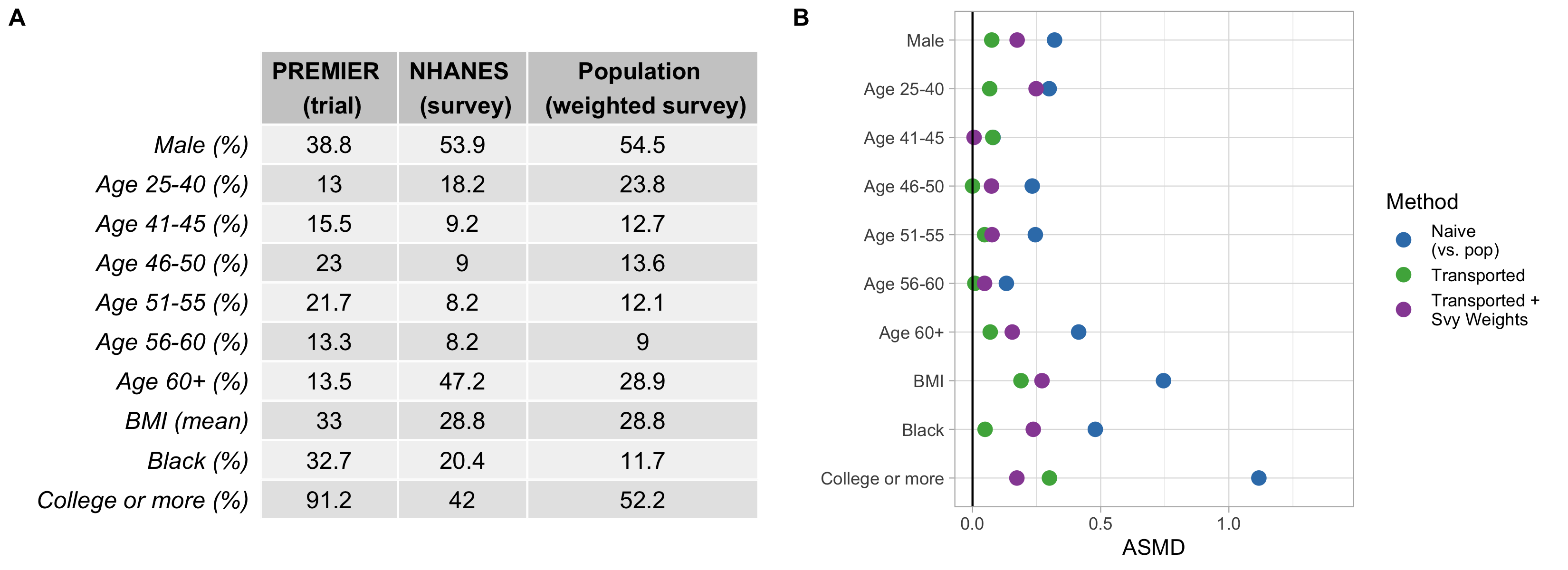}}
\caption[A) Covariate Distributions in PREMIER (trial) and NHANES (survey sample), along with the weighted NHANES sample (target population). B) Absolute standardized mean difference (ASMD) of covariates between the trial and target population.]{A) Covariate Distributions in PREMIER (trial) and NHANES (survey sample), along with the weighted NHANES sample (target population). B) Absolute standardized mean difference (ASMD) of covariates between the trial and target population. Points in blue reflect covariate differences between the raw trial sample and the weighted survey sample (i.e. the target population demographics). Points in green show the differences between the transport-weighted trial and survey sample. Points in purple show the differences between the transport-weighted trial and population (where the trial is weighted to be more similar to the target population).}
\label{fig:covariatesnutrition}
\end{figure}

Figure \ref{fig:covariatesnutrition}A shows the covariate distributions in the trial and survey samples, as well as in the weighted survey sample (indicating the target population of interest). Observe that while some variables, such as sex and BMI, are distributed quite similarly between the unweighted and weighted NHANES samples, other variables, such as race, age and education, differ a fair amount between the two. These differences show the NHANES survey sampling methodology, and how the true population characteristics may differ from the raw analytic sample. If we generalize to the NHANES survey sample (i.e. fit the transportability estimator ignoring the survey sampling weights), we would be generalizing to a population that is younger, less educated, and more racially diverse than our true target population of interest. Figure \ref{fig:covariatesnutrition}B shows the covariate balance between the trial and target population before and after transport weighting. Note that weighting the trial to resemble either the NHANES sample or the target population results in better covariate balance; however, only the latter is truly relevant to our interests.

\begin{figure}
 \centerline{\includegraphics[width=100mm]{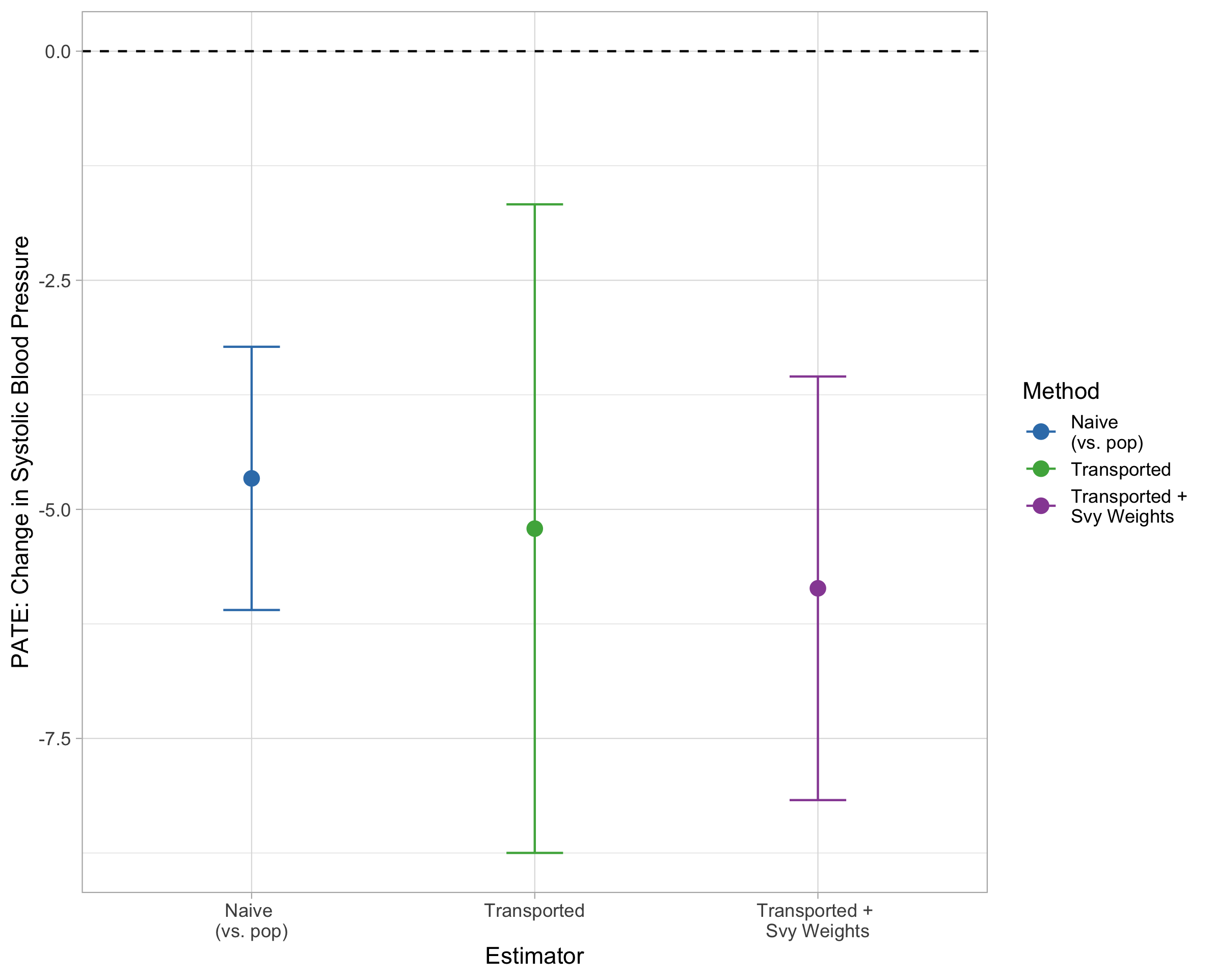}}
\caption[Blood pressure reduction PATE estimates by transportability method.]{Blood pressure reduction PATE estimates by transportability method. Points in blue reflect the naive PATE estimate, points in green show the transported PATE estimate ignoring survey weights. Points in purple show the survey-weighted transportability estimate.}
\label{fig:resultsnutrition}
\end{figure}

The effect of the lifestyle intervention on change in SBP is shown in Figure \ref{fig:resultsnutrition}, with the naive trial estimate on the left, the transported estimate in the middle, and the transported estimate using survey weights on the right. The naive trial estimate of -4.66 and 95\% confidence interval of (-6.10, -3.23) indicate a positive effect of the lifestyle intervention recommendations in lowering systolic blood pressure among study participants, as originally reported in the trial findings. In this example, there are no substantial differences between the naive estimates and the transported estimates, nor between the two transported estimates (ignoring vs. incorporating the survey weights). Note, though, that both weighted estimators have larger standard errors. Given the consistent estimates, these generalized findings provide further evidence to support the original trial's claims, that PREMIER's results should be used to influence blood pressure management policies related to persons with above-optimal blood pressure in the United States.

%\begin{table}[]
%\centering
%\caption{Covariate Distributions in PREMIER (trial) and NHANES (survey sample), along with the weighted NHANES sample (target population). Note how the use of survey weights changes the covariate distribution in NHANES, reflecting the importance of using the survey weights when transporting from the trial to the true target population demographics. \label{tab:covtabnutrition}}
%\begin{tabular}{lccc}
%\toprule
%  & \thead{PREMIER\\ (trial)} & \thead{NHANES \\ (survey)}& \thead{Population \\ (weighted survey)}\\
%\midrule
%\rowcolor{gray!6}  Male & 0.39 & 0.54 & 0.54\\
%\addlinespace[0.3em]
%\multicolumn{4}{l}{\textbf{Age}}\\
%\hspace{1em}25-40 & 0.13 & 0.18 & 0.24\\
%\rowcolor{gray!6}  \hspace{1em}41-45 & 0.15 & 0.09 & 0.13\\
%\hspace{1em}46-50 & 0.23 & 0.09 & 0.14\\
%\rowcolor{gray!6}  \hspace{1em}51-55 & 0.22 & 0.08 & 0.12\\
%\hspace{1em}56-60 & 0.13 & 0.08 & 0.09\\
%\rowcolor{gray!6}  \hspace{1em}60+ & 0.13 & 0.47 & 0.29\\
%BMI & 32.97 & 28.79 & 28.81\\
%\rowcolor{gray!6}  Black & 0.33 & 0.20 & 0.12\\
%Education (College or more) & 0.91 & 0.42 & 0.52\\
%\bottomrule
%\end{tabular}
%\end{table}

\subsection{Web-Based Intervention for Treating Substance Use Disorders}
\label{s:datasud}
We now turn to our second illustrative example using a trial from the Clinical Trials Network (CTN), a publicly available data repository for substance use-related RCTs funded by the National Institute of Drug Abuse (NIDA). The trial of interest, CTN-0044, evaluated the effectiveness of Therapeutic Education System (TES), a web-based behavioral intervention including motivational incentives, as a supplement to SUD treatment. A total of 507 individuals in treatment for SUDs were randomized to either treatment as usual or treatment plus TES, and the reported trial results suggested that TES successfully reduced treatment dropout and improved upon abstinence \citep{campbell2014internet}. Our outcome of interest is a binary indicator of drug and alcohol abstinence in the last week of the study. 

We generalize these findings from CTN-0044 to a population of US adults seeking treatment for substance use disorders using NSDUH, a survey on drug use in the United States. In its sampling design, NSDUH systematically over-samples adults over the age of 26 in order to better estimate drug use and mental health issues in the US. This suggests that the raw NSDUH survey sample is likely not reflective of the target population on key demographics. We subset the NSDUH sample to individuals over the age of 18 who have reported any illicit drug use in the past 30 days in order to best reflect our target population of interest. The resulting NSDUH sample has 5645 participants representing a target population of around 20 million people.

\begin{figure}
 \centerline{\includegraphics[width=145mm]{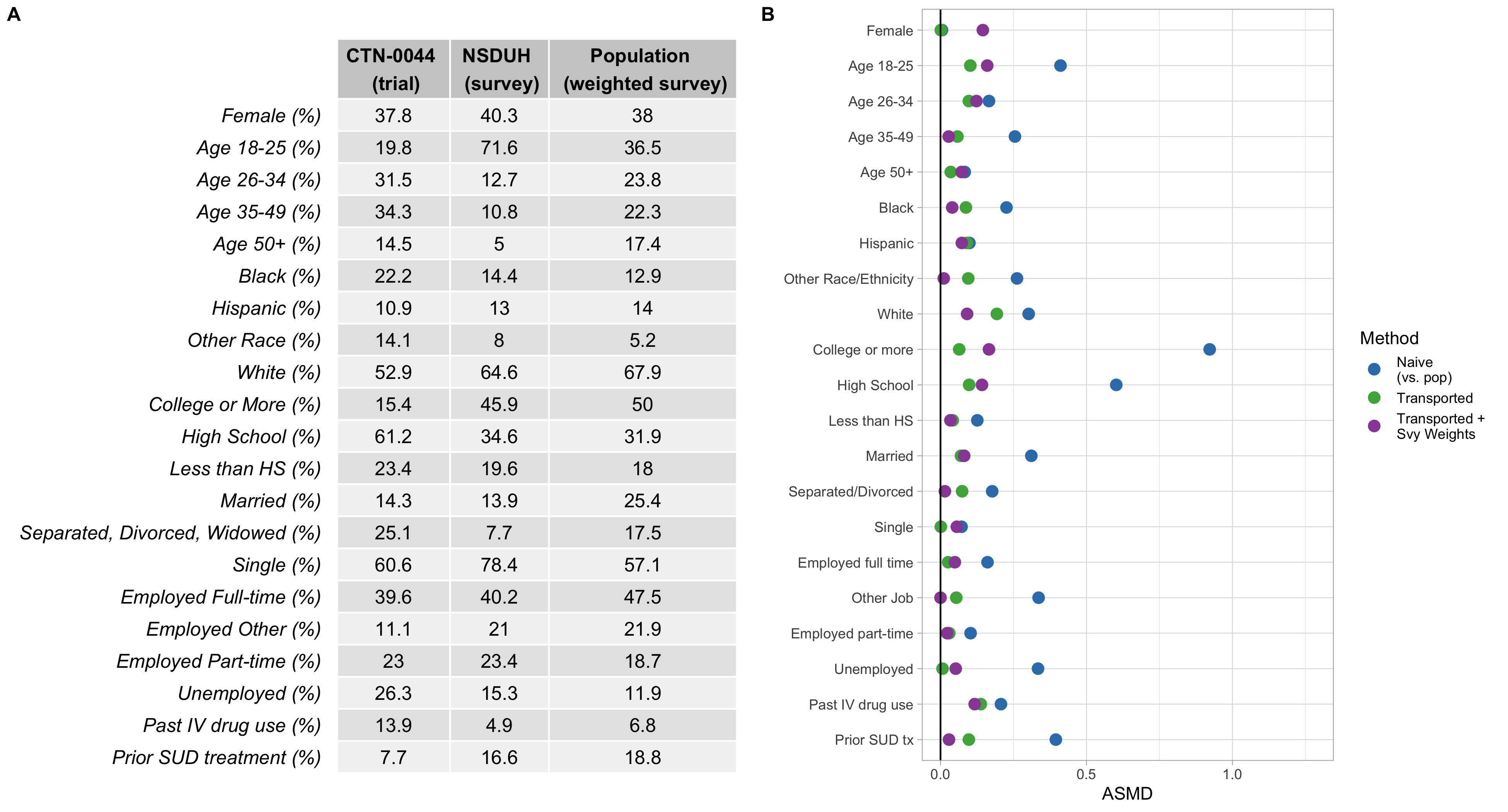}}
\caption[A) Covariate Distributions in CTN-0044 (trial) and NSDUH (survey sample), along with the weighted NSDUH sample (target population). B) Absolute standardized mean difference (ASMD) of covariates between the trial and target population.]{A) Covariate Distributions in CTN-0044 (trial) and NSDUH (survey sample), along with the weighted NSDUH sample (target population). B) Absolute standardized mean difference (ASMD) of covariates between the trial and target population. Points in blue reflect covariate differences between the raw trial sample and the weighted survey sample (i.e. the target population demographics). Points in green show the differences between the transport-weighted trial and survey sample. Points in purple show the differences between the transport-weighted trial and population (where the trial is weighted to be more similar to the target population).}
\label{fig:covariatessud}
\end{figure}

The distribution of covariates across the trial and survey samples are shown in Figure \ref{fig:covariatessud}A. Note that, pre-transport-weighting, there are substantial differences between the trial and raw survey samples with respect to age, though when the survey sample is weighted to the target population using the survey weights, these age differences decrease. Other variables like race, education and prior substance use treatment are actually \emph{more} different between the trial and target population than they are between the trial and unweighted NSDUH sample. This further highlights the importance of incorporating the survey weights in order to make inferences on the true target population of interest when transporting. Figure \ref{fig:covariatessud}B shows the covariate balance between the trial sample and (survey-weighted) target population before and after weighting. Points in green show covariate balance when the trial is weighted to the raw survey sample, while points in purple show covariate balance when the trial is weighted to the target population (the survey-weighted survey sample). Overall, both weighting methods yield better balance (and therefore better resemblance) between the trial and the population, though it should be noted again that only the points in purple reflect when the trial is weighted to resemble the true target population (i.e. the survey weights are used in the sample membership model). 

\begin{figure}
 \centerline{\includegraphics[width=120mm]{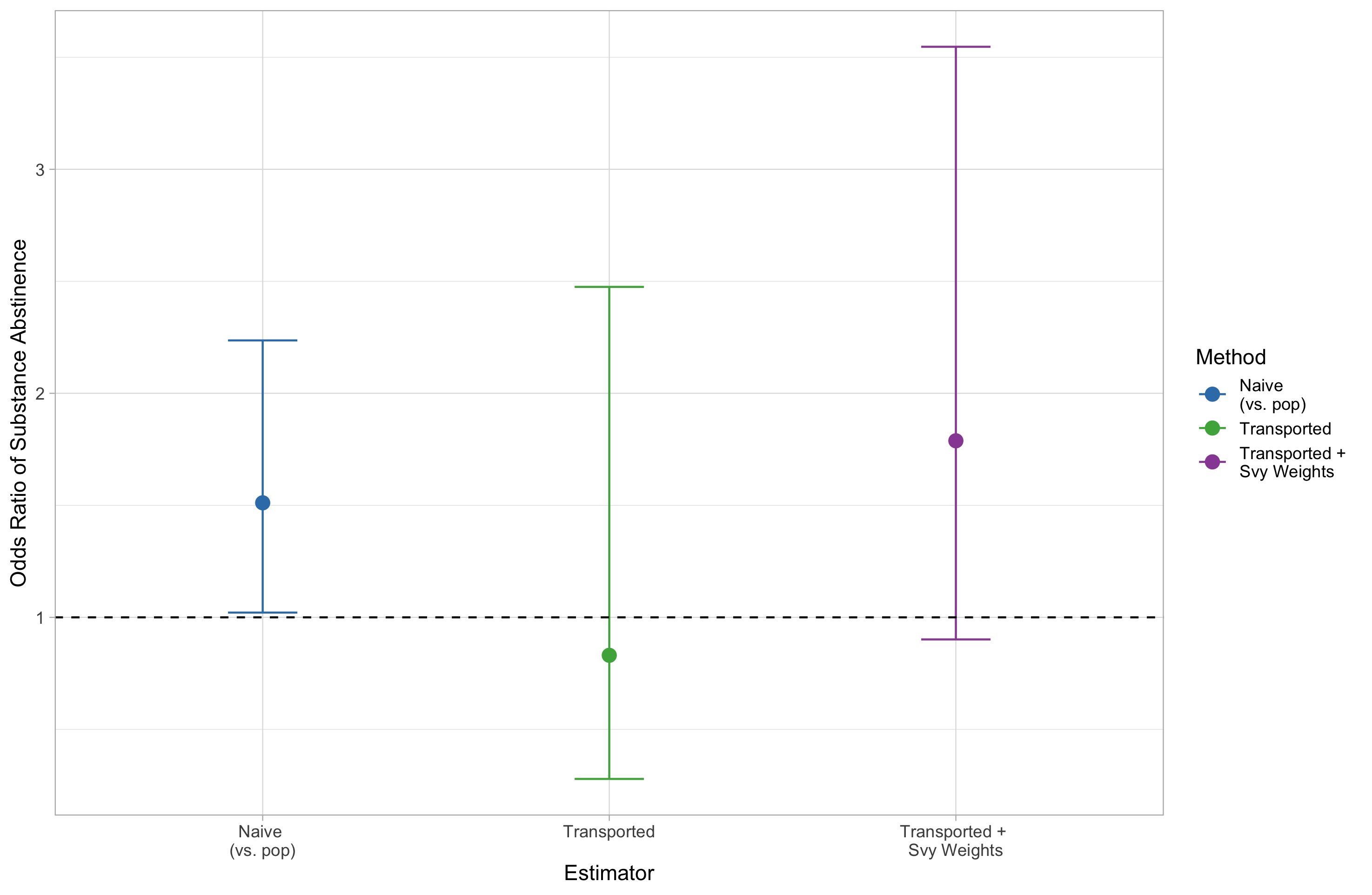}}
\caption[Substance abstinence PATE estimates by transportability method.]{Substance abstinence PATE estimates by transportability method. Points in blue reflect the naive PATE estimate, points in green show the transported PATE estimate ignoring survey weights. Points in purple show the survey-weighted transportability estimate.}
\label{fig:resultssud}
\end{figure}

Figure \ref{fig:resultssud} depicts the three PATE estimates, or the odds ratio of substance abstinence. As reported in the original trial, the naive estimate is statistically significant, with an odds ratio of 1.5 and 95\% confidence interval of (1.02, 2.24), suggesting that TES was effective in increasing substance abstinence in the trial. However, when this estimate is transported to the NSDUH sample (middle, green), this point estimate drops to around 0.8 and the confidence interval width increases (0.28, 2.48). While the lower odds ratio may suggest qualitative differences in TES' effectiveness, the transported estimate indicates no significant difference in abstinence rates between the two treatment arms in the NSDUH sample. When the survey weights are included in the sample membership model, and the estimate therefore generalized to the target population of interest, the wide confidence interval of (0.90, 3.55) indicates a similar not-significant conclusion, though the point estimate of 1.79 more closely mirrors what was estimated in the original trial. This example highlights that if the survey weights are left out when making generalizations, different qualitative conclusions may be reached.

\section{Conclusion}
\label{s:conclusion}
Existing methods for improving RCT generalizability with propensity score-type weights make an implicit assumption about the population data: that they are either 1) a simple random sample drawn from the true target population, or 2) the \emph{complete} finite target population. When transporting trial findings to a population dataset that come from a complex survey, this assumption no longer holds. Our work demonstrates that it is crucial to incorporate the survey weights from the complex survey population data in order to obtain the best estimate of the PATE with these methods. Omitting the survey weights can be thought of as generalizing to an entirely different population, one that has the demographics of the survey sample rather than the target population of interest. While the demographic differences between a survey sample and its intended target population may not be that large for some analytic survey datasets, it can be particularly noticeable for others where great amounts of over- or under-sampling of certain groups are implemented.

Our work has shown that fitting a sample membership model weighted by survey weights can only improve upon our ability to draw population-level inferences from RCTs, and that failing to do so (i.e. using standard transportability weights alone) may actually result in \emph{more} biased estimates. Given that complex survey data often come ready for use with a variable containing the necessary survey weights, implementing this approach does not require specifying any additional models other than those needed for the standard transportability weighting methods. Still, there are still a few limitations to this work. First, as noted earlier in this paper, we can obtain an unbiased estimate the PATE when we assume that \emph{all} covariates impacting survey selection, trial selection, \emph{and} treatment effect heterogeneity are fully observed and accounted for in both datasets. In practice, certain variables used to construct the survey weights may not be publicly available at the individual-level in the survey sample. While we demonstrated the performance of these methods when we use a correlated proxy for one such variable, it is also conceivable that certain key covariates may be unobserved in one or both datasets completely. Further research is needed to extend upon sensitivity analyses for partially and fully unobserved treatment effect modifiers, particularly when the population data come from a complex survey. Second, while we explored the benefit of double-bootstrapping methods for variance estimation, there may be additional concerns over uncertainty introduced by using a small survey sample that represents a huge target population. Additional research is warranted to assess the impact of the proportion of population sampled on estimate variability. Finally, the propensity score-type weighting method explored in this paper is only one post-hoc statistical approach for estimating population effects from RCTs. Outcome-model-based approaches have also been shown beneficial, where a model is fit using trial data, and predictions are generated under treatment and control conditions in the target population data. Future work should build upon such methods when using complex survey population data as well. Nevertheless, our two-stage weighting method will ultimately allow researchers to draw more accurate inferences from trials to be used in policy formation and population-level decision making.

\clearpage

\section{Appendix}
\label{s:appendix}

\subsection{Derivation of Population Estimand E[Y(a)] for single binary X}

\begin{align*}
E[Y(a)] &= \sum_{x} E[Y(a)|X = x]P(X = x) && \text{Total expectation}\\
&= \sum_{x} E[Y(a)|X = x,S=1]P(X = x) && S \indep Y(a)|X\\
&= \sum_{x} E[Y(a)|A=a,X = x,S=1] P(X = x) && A \indep Y(a)|X, S = 1 \\
&= \sum_{x} E[Y|A=a,X = x,S=1] P(X = x) && \text{ Consistency }\\
&= \sum_{x} E[Y|A=a,X = x,S=1] P(X = x|S=1) \\
&\hspace{2cm} \times \frac{P(S = 1)}{P(S = 1|X = x)} && \text{ Bayes thm }\\
E[Y(a)]  &= \sum_{x} E[Y|A=a,X = x,S=1]P(X=x|S=1) \\
&\hspace{2cm} \times \Bigg( \frac{P(S = 2 | X = x)P(S = 1)}{P(S = 1|X = x)P(S = 2)} \Bigg)\\
&\hspace{2cm} \times \Bigg(\frac{P(S = 2)}{P(S = 2 | X = x)} \Bigg) && \text{ multiplying by 1}
\end{align*}

\subsection{Double Bootstrap}
\begin{figure}
 \centerline{\includegraphics[width=145mm]{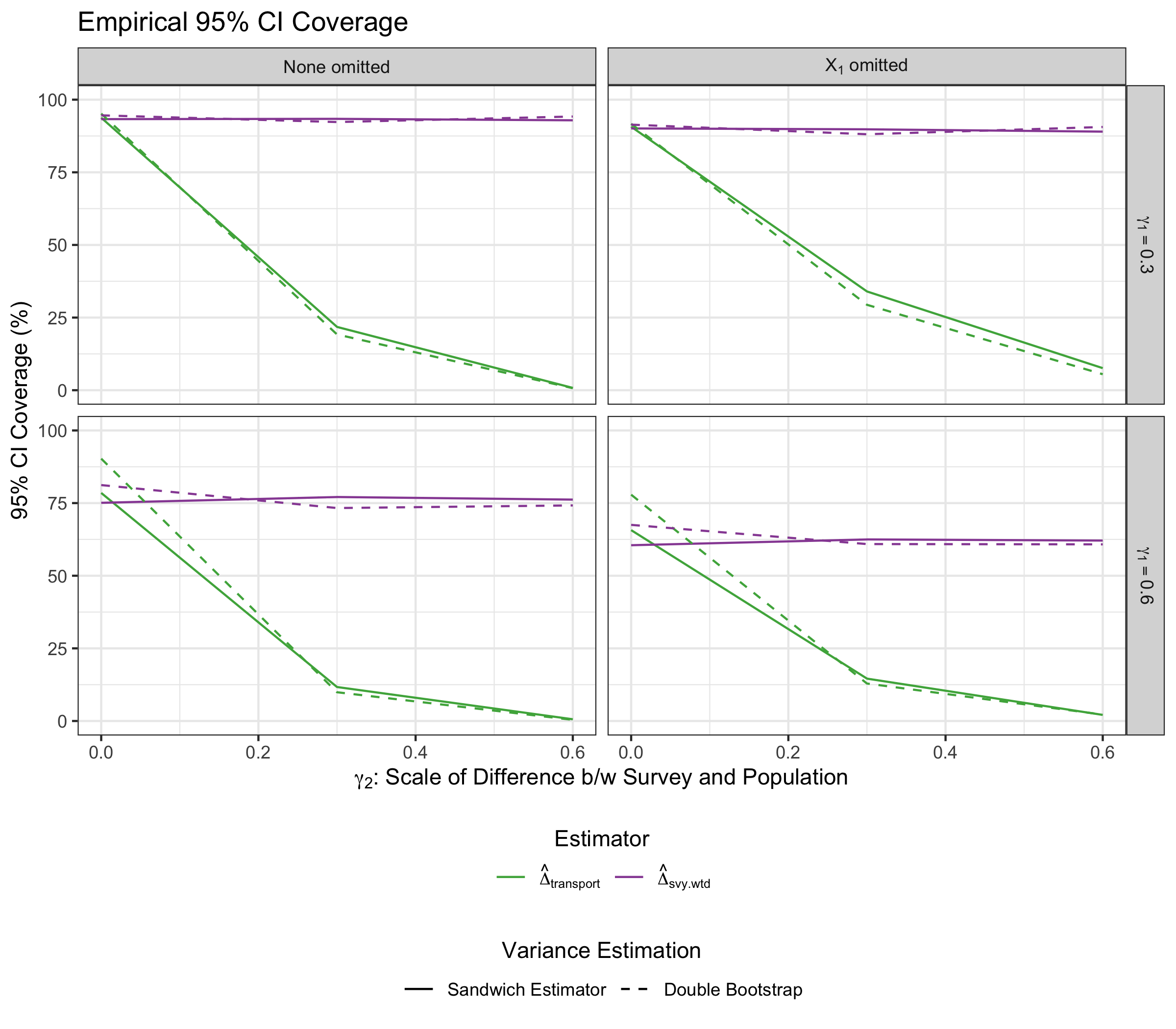}}
\caption{Empirical coverage of the transportability estimators using the double bootstrap approach to estimate the variance.}
\label{fig:coveragebstrap}
\end{figure}

In order to account for uncertainty in the survey when using it for generalizations, we propose using a double-bootstrapping approach to estimate the variability of the PATE estimates. Similarly to how a bootstrap involves sampling with replacement many times and looking at the distribution of estimates across bootstrap runs, we sample both the trial \emph{and} the survey with replacement in each bootstrap run. Within each bootstrap iteration, we re-sample the trial with replacement (sample size equal to that of the trial). We also re-sample the survey using a stratified approach described by \citet{valliant2013practical}. We define survey strata by deciles of the survey weights. For stratum $h$ with sample size $n_h$, we sample with replacement $m_h = n_h - 1$ subjects. We adjust the survey weight $d_k$ of subject $k$ to equal $$d_k^{*} = d_k \frac{n_h}{n_h - 1} m_{hi}^{*} $$ where $m_{hi}^{*}$ is the number of times subject $k$ is sampled for that given bootstrap run. Therefore, if the subject is selected once, their new weight is equal to $d_k \frac{n_h}{n_h - 1}$. If they are selected twice, their new weight is equal to $d_k \frac{2n_h}{n_h - 1}$, and so forth. Figure \ref{fig:coveragebstrap} compares the empirical 95\% coverage of the transported estimators using this double bootstrap approach on a subset of the simulation scenarios to the standard sandwich variance estimator used for Figure \ref{fig:coverageresults}. Note that the results across the different approaches are quite similar, though the double bootstrap yields slightly better coverage when the trial differs more from the target population (bottom row).

%\clearpage

\end{document}